\documentclass[twocolumn,prl,aps,superscriptaddress,showpacs]{revtex4}
\usepackage{mathrsfs}
\usepackage{graphicx}
\begin{document}
\title {Giant vortex and Skyrmion in a rotating two-species Bose-Einstein condensate}
\author{Shi-Jie Yang\footnote{Electronic address: yangshijie@tsinghua.org.cn}}
\author{Quan-Sheng Wu}
\author{Sheng-Nan Zhang}
\author{Shiping Feng}
\affiliation{Department of Physics, Beijing Normal University,
Beijing 100875, China}
\begin{abstract}
Numerical simulations are performed for a rotating two-species Bose
condensate confined by a harmonic potential. The particle numbers of
each species are unequal. When the rotational speed exceeds a
critical value, the majority species reside in the center of the
potential while the minority species is pushed out to the outskirts,
forming a giant vortex hole to contain the majority species. A novel
annular Skyrmion forms at the interface of the two species.
\end{abstract}
 \maketitle

\section{introduction}
Quantum coherence enables intriguing phenomena in Bose-Einstein
condensates (BEC) as quantized vorticity, which has been verified in
many experiments. When a trapped BEC is driven to rotate, singly
quantized vortices form. Faster rotation generates more vortices
which finally condense into a lattice. Other methods, such as the
phase imprint technique, are also used to create vortex in BECs.
D.R. Scherer et al\cite{Scherer} carried out an experiment to
implement vortices by the interference of three independent trapped
BECs.

There are many efforts to create a multiply quantized vortex or
giant vortex in the BECs. However, a giant vortex is energetically
disfavored in a harmonic trap and is not expected to persist if
created in a rotating condensate. Some authors tried to overcome
this dissociation instability by employing an external repulsive
pinning potential\cite{Simula} or a steeper trap that allows
rotation at frequency exceeding the centrifugal limit of the
harmonic trap.\cite{Fetter,Lundh,Kasamatsu,Fischer,Cozzini} Multiply
quantized vortices are also created topologically by deploying the
spin degrees of freedom in optical traps.\cite{Isoshima} In a recent
experiment, P. Engels et al shone a near resonant laser beam through
a rapidly rotating harmonically trapped BEC to produce a density
hole encircled by a high number of vorticity.\cite{Engels,Simula2}

In this work, we produce a giant vortex in a harmonic trap by
introducing a second species of BEC. There is a population imbalance
($N_1>N_2$) in the two-species BEC and the inter-species scattering
length is larger than the intra-species length. We consider a
quasi-2D repulsive system by numerically integrating the
Gross-Pitaevskii (GP) equation. S. Bargi et al\cite{Bargi} have
studied the same setup by numerical diagonlization and variational
mean-field approxiamtion. In the limit of $N_1\gg N_2$, the density
profile of the majority species is approximately the same as if the
minority species is absent. On the other hand, the minority species
feels a trap formed by the external magnetic trap plus the hump in
the center which is generated by the majority species. Hence the
effective potential acted on the second species behaviors like a
'Mexican hat'. The minority species is pushed to the rims, which
makes for the formation of a hole. When this system is set to
rotate, a series of singly quantized vortices are created. Large
centrifugal force induced by faster rotation subsequently generates
a giant hole in the minority species to contain the majority
species. A number of phase defects congregated in the central area
reveals that this giant hole is a multiply quantized vorticity. This
becomes possible because of the low density in the hole costing
little energy for vortices to merge into the giant vortex. By
describing the system in terms of pseudo-spin density parameters, it
is found that associated to the giant vortex is an annular Skyrmion.
Although there are many works to deal with Skyrmions resulted from
singly quantized
vortices,\cite{Ho,Mizushima,Isoshima2,Kasamatsu2,Mueller} the
present multiply quantized vortex creates a Skyrmion with novel and
more complex topological structure. The topological charge density
exhibits a ring distribution.

\section{formulism}
In weak interaction limit, quantum dynamics of the BEC is governed
by the nonlinear Gross-Pitaevskii equation. We consider a 2D system
that subjects to rotation $\mathbf{\Omega}=\Omega \hat{z}$. In a
frame rotating with frequency $\Omega$ around the $z$ axis, the
dynamics of a condensate wavefunction $\Psi=(\psi_1,\psi_2)^T$ is
described by the following normalized equation
\begin{equation}
i\frac{\partial \psi_i}{\partial t}=[-\frac{1}{2}\nabla^2
+V(\vec{r})+\sum_{j=1,2} g_{ij}|\psi_j|^2-\Omega \hat{L}_z] \psi_i,
\label{GP}
\end{equation}
where $g_{ij},(i,j=1,2)$ are the nonlinear coupling constants which
are expressed by $g_{ij}=4\pi N_j a_{ij}^s$, with $a_{ij}^s$ the
inter ($i\neq j$) or intra ($i=j$) species s-wave scattering
lengthes. $V(\vec r)=\frac{1}{2}\omega^2 (x^2+y^2)$ is the harmonic
trapping well. We assume that the two species of BEC are trapped by
the same external potential with unequal particle numbers $N_1>N_2$
and all the nonlinear coupling constants are positive.

We first analyze the problem in the limit of $N_1\gg N_2$. The
distribution of the majority species is hardly affected by minority
species. Hence we can approximate its profile with the Thomas-Fermi
(TF) distribution, $n_1(\vec r)=|\psi_1(\vec
r)|^2=[\mu_1-\frac{1}{2}(\omega^2-\Omega^2) r^2]/g_{11}$ for
$n_1(\vec r)>0$ and $n_1(\vec r)=0$ otherwise. $\mu_i$ is the
chemical potential. The term associated with the rotation frequency
$\Omega$ is the centrifugal potential. We consider the case for
$\Omega<\omega$. The radius $R_1$ of the condensate is evaluated by
finding where the density goes to zero. If $\mu_1>0$ then there is a
single solution with $R_1=\sqrt{2\mu_1/(\omega^2-\Omega^2)}$. The
chemical potential is determined by normalization condition $\int
n_1(\vec r)d\vec{r}=1$, which gives rise to
$\mu_1=\sqrt{g_{11}(\omega^2-\Omega^2)/\pi}$. Because of the
strongly repulsive interspecies coupling, the minority is pushed out
to the rim which feels an effective potential produced by the
external trap combining with the repulsive hump of the majority,
$V_{eff}(\vec r)=V(r)+g_{21}n_1(\vec r)$, which looks like a
'Mexican hat'. The minority condensate prefers to reside in annular
notches of the effective potential and subsequently forms a huge
hole in the central area to contain the majority condensate.

The Thomas-Fermi profile of the minority condensate is
\begin{equation}
n_2(\vec r)=|\psi_2(\vec r)|^2=(\mu_2-V_{eff}(\vec r)
+\frac{1}{2}\Omega^2 r^2)/g_{22},
\end{equation}
for $n_2(\vec r)>0$ and $n_2(\vec r)=0$ otherwise. The outer and
inner radii of the ring-shape condensate are
\begin{eqnarray}
R_2^{+}&=&\sqrt{\frac{2\mu_2}{(\omega^2-\Omega^2)}} \\\nonumber
R_2^{-}&=&\sqrt{\frac{2(g_{21}/g_{11}\mu_1-\mu_2)}{(g_{21}/g_{11}-1)(\omega^2-\Omega^2)}}.
\end{eqnarray}
The corresponding chemical potential is given by
\begin{equation}
\mu_2=\mu_1-\frac{g_{11}}{g_{21}}\sqrt{\frac{g_{22}}{\pi}(\frac{g_{21}}{g_{11}}-1)(\omega^2-\Omega^2)},
\end{equation}
which requires $g_{21}>g_{11}$. In this case, the two species of BEC
is imiscible.

\begin{figure}[tbh]
\includegraphics[width=8cm]{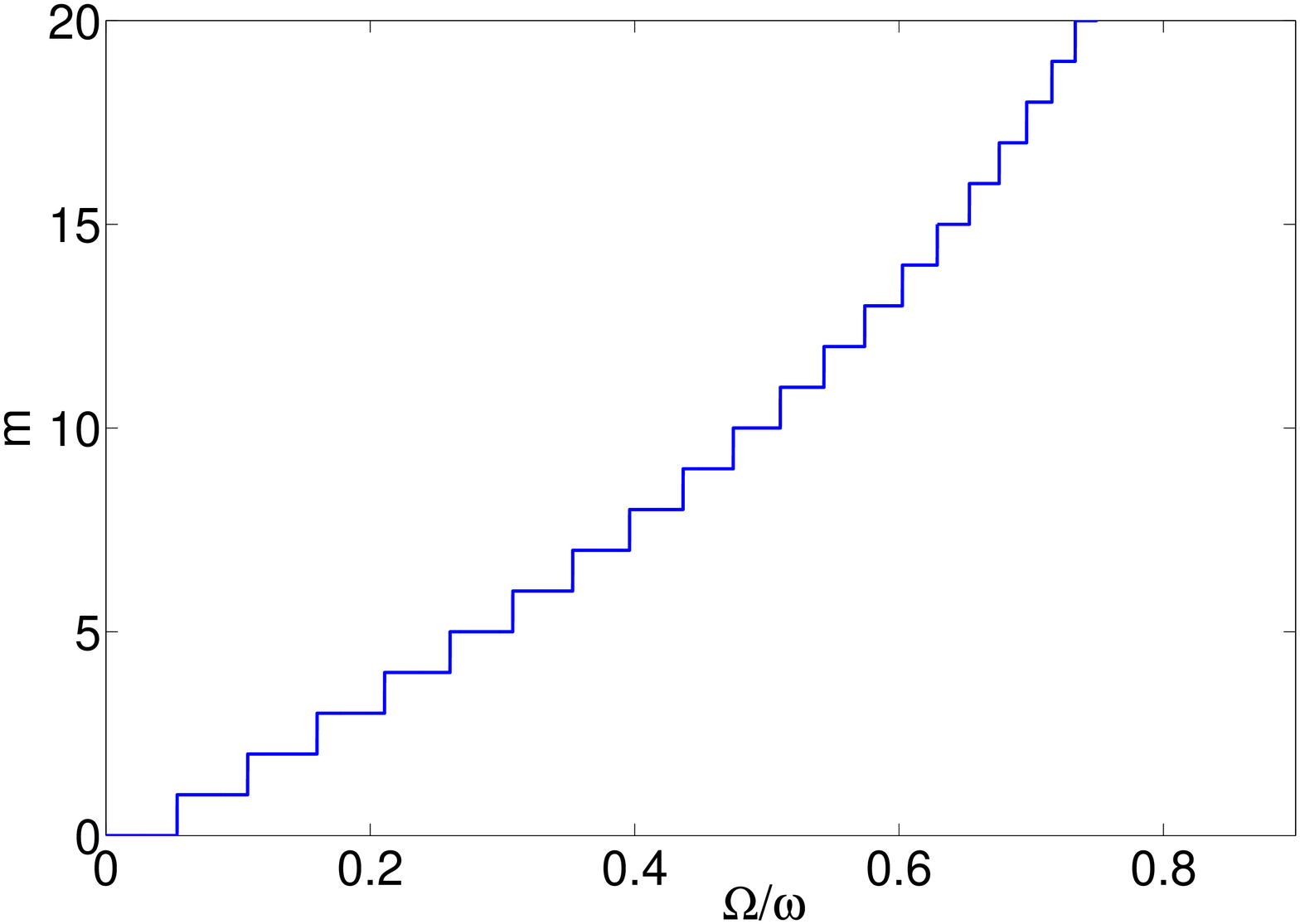}
\caption{The rotational frequency dependence of total vorticity from
formula [\ref{vorticity}].}
\end{figure}

Suppose there is a vortex in the center with circulation
$\Gamma=2\pi m$. The quantization condition requires the vorticity
of integer values $m=0,1,2,\cdots$. The energy of the minority
condensate in the TF approximation is expressed as,
\begin{equation}
E=\int d\vec r [V_{eff}+\frac{m^2}{2r^2}+\frac{1}{2}g_{22}n_2(\vec
r)] n_2(\vec r). \label{energy}
\end{equation}

In the above analysis we have omitted the contribution from singly
quantized vortices which may form a vortex lattice. The energy in a
frame rotating with angular velocity $\Omega$ is related to that in
the laboratory frame by $E'(m)=E(m)-m\Omega$. For a given rotation
frequency $\Omega$, this energy is to be minimized with respect to
the parameter $m$ to obtain the vorticity of the giant vortex. By
treating the parameter $m$ as continuous variable, the final result
is
\begin{widetext}
\begin{equation}
m=\frac{g_{22}\Omega}{2\pi
\{\mu_2\ln(\frac{R_2^+}{R_2^-})-\frac{g_{22}}{g_{11}}\mu_1\ln(\frac{R_1}{R_2^-})+\frac{1}{4}(\omega^2-\Omega^2)
[\frac{g_{21}}{g_{11}}((R_1)^2-(R_2^-)^2)-((R_2^+)^2-(R_2^-)^2)]\}}.\label{vorticity}
\end{equation}
\end{widetext}

Figure 1 depicts the vorticity dependence on the rotational speed
$\Omega$. It should be noted that the total vorticity also includes
the singly quantized vortices. Hence Eq.[\ref{vorticity}] is a
schematic formula for the relation between the vorticity and the
rotational speed.

\section{numerical simulations}
The numerical simulations is carried out by solving the
norm-preserving imaginary time propagation of Eq. [\ref{GP}]. The
initial wavefunction adopts the Thomas-Fermi approximation. We adopt
the time-splitting Fourier pseudospectral method developed by Bao et
al\cite{Bao} to compute the partial differential equation [\ref{GP}]
in a region $(x,y)\in [-8,8]\times [8,8]$ with a refined grid of
$80\times 80$ nodes, which is sufficient to achieve grid
independence. The particle number ratio is fixed at $N_1/N_2=2$ and
the nonlinear coupling parameters are chosen as $g_{11}=100$,
$g_{22}=550$, $g_{12}=200$, and $g_{21}=400$.

\begin{figure}[t]
\includegraphics[width=8cm]{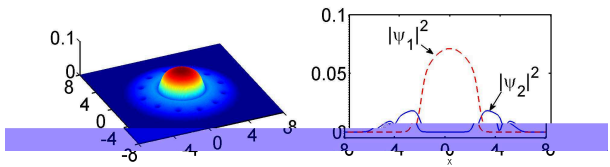}
\caption{Density profiles of the two-species BEC for rotational
speed $\Omega/\omega=0.85$. The ratio of particle numbers is
$N1/N2=4$. In the minority species there forms a ring of singly
quantized vortices plus a huge hole in the center.}
\end{figure}

Figure 2 shows the density profile of both species of condensates.
As we have expected, the majority condensate resides in the central
area of the harmonic well while pushes the minority condensate to
the outskirts, enabling the latter to acquire more angular momentum
to circulate around it. In the mean time, the majority condensate
becomes more compact by the inward force implemented by the minority
condensate. As the rotating speed increases, a lot of phase defects
of the minority condensate enter the center area of the harmonic
trap and merge into a giant vortex. Here the giant vortex implies
that several phase defects are contained in a single density hole.
Figure 3 exhibits that the circulating movement of the minority
condensate forms a giant vortex with winding number $m=8$. In
addition, a sequence of singly quantized vortex forms along the
ring-shape minority condensate.

\begin{figure}[tbh]
\includegraphics[width=8cm]{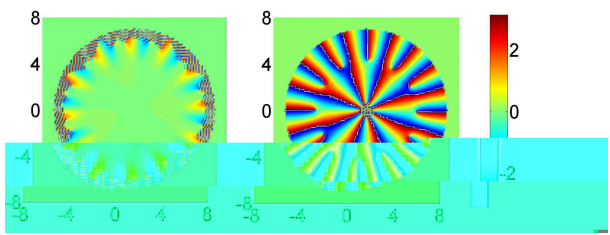}
\caption{Phase plots for (a) the majority species and (b) the
minority species. The parameters are the same as in Fig.1. Panel (b)
indicates that the huge hole is a multiply quantized vortex with a
winding number $m=8$.}
\end{figure}

To further reveal the explicit structure of the giant vortex, we
introduce a normalized complex valued spinor $\chi=[\chi_1(\vec
r),\chi_2(\vec r)]^T$ by decomposing the wavefunction
$\psi_i=\sqrt{\rho_T(\vec r)}\chi_i(\vec r)$, where $\rho_T$ is the
total density. They satisfy $|\chi_1|^2+|\chi_2|^2=1$. The
pseudo-spin density is defined as $\vec S=\chi^\dagger (\vec r)\vec
\sigma \chi (\vec r)$, where $\vec \sigma$ is the Pauli matrix.
Obviously, the modulus of the total spin is $|\vec S|=1$. The
pseudo-spin texture corresponding to the giant vortex is shown in
Fig. 4. It is shown that the $\vec S_x$ or $\vec S_y$ exhibits an
$m$-fold symmetrical annular necklace of radius $R$, which is
clearly related to the quantized circulation of the giant vortex. As
$\vec S_x=2|\chi_1||\chi_2|\cos (\theta_1-\theta_2)$, where
$\theta_i(\vec r)$ are the phases of the condensate wavefunctions,
and from Fig.3 the phase of the majority condensate is approximately
constant, we know that $\vec S_x$ has $m$-fold symmetry. However, if
the rotation frequency $\Omega$ increases further, vortices are
created in the majority condensate and $\theta_1$ can not be viewed
as constant, then the $m$-fold symmetry should be broken. Fig.4(b)
is the $\vec S_z$ distribution which indicates that the pseudo-spin
points up at the center and points down on the edge which results
from $\chi_2\approx 0$ for $r\lesssim R$ and $\chi_1\approx 0$ for
$r\gtrsim R$. At the interface of the two condensates, the system
twists its pseudo-spinor order parameter, as shown in the projected
vectorial plot of ($\vec S_x,\vec S_y$) in Fig.4(c). It reveals a
rather complex pseudo-spin texture which may be called a giant
Skyrmion.
\begin{figure}[tbh]
\includegraphics[width=8cm]{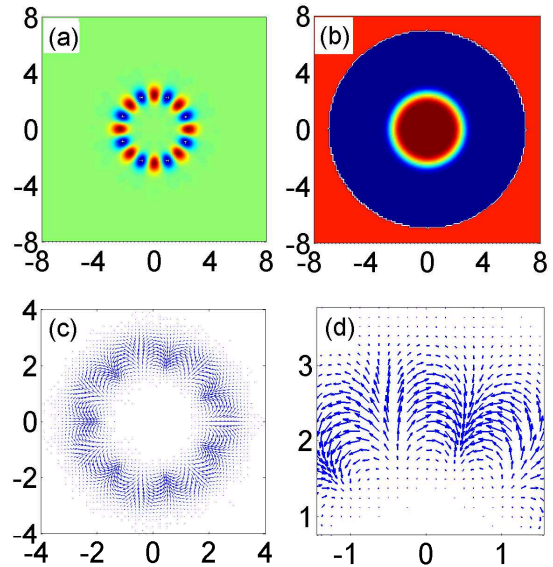}
\caption{Pseudo-spin density distribution for (a) $S_x$, (b) $S_z$.
Panel (c) is the vectorial plot of $(S_x,S_y)$, which exhibits an
annular Skyrmion structure. (d) is an amplified local part of (c).}
\end{figure}

The pseudo-spin texture in Fig. 4 reveals a concrete $k$-fold
symmetry. At first sight, it seems curious that $k=m-1$ instead of
$k=m$. It is because the pseudo-spin contributes is an additional
rotational angle $\Delta\phi=2\pi/k$ that relates to the spatial
rotation. Taking the spins to rotating about the $\hat z$ axis, the
symmetry is formally described by
\begin{equation}
S_i(R(2\pi/k)\vec r)=\mathcal{R}_{ij}(2\pi+\Delta\phi) S_j (\vec r),
\label{rotate}
\end{equation}
where
\begin{equation}
R(\phi)=\left(\begin{array}{cc}
   \cos \phi & \sin \phi \\
  -\sin \phi & \cos \phi
\end{array}\right)
\end{equation}
represents a spatial rotation about the $\hat z$ axis by an angle
$\phi$, and $\mathcal{R}_{ij}(\phi)$ ($i,j=x,y$) represents a
pseudo-spin rotation about the $\hat z$ axis by an angle $\phi$. The
projected pseudo-spin vector increases an angle of $2\pi m$ when it
runs along a route that encloses the origin.

The left panel of Fig. 5 shows the effective velocity field that is
defined as
\begin{equation}
\vec v_{eff}(\vec r)=(\vec j_1(\vec r)+\vec j_2(\vec
r))/\rho_{T}(\vec r),
\label{velocity}
\end{equation}
with $\vec j_i=\frac{1}{2i}(\psi_i^*\bigtriangledown
\psi_i-\psi_i\bigtriangledown \psi_i^*)$, $(i=1,2)$ the partial
current density. Besides a sequence of singly quantized vortices,
the whole minority condensate circulates around the majority
condensate with a multiply quantized circulation $\Gamma=8\times
2\pi$. Contrary to the conventional vortex in a single species
condensate, $|\vec v_{eff}|$ vanishes at the center which makes a
coreless vortex without a density dip in the total density. The
largest curl of the velocity field lies on a ring of finite radius.

\begin{figure}[tbh]
\includegraphics[width=8cm]{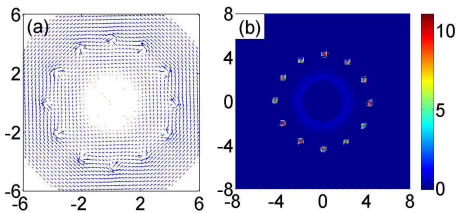}
\caption{(Color online) (a) The vectorial plot of the effective
velocity $\vec v_{eff}$ defined by Eq.[\ref{velocity}]; (b) The
topological charge density $q(\vec r)$ distribution.}
\end{figure}

The topological charge density is defined as
\begin{equation}
q(\vec r)=\frac{1}{8\pi}\epsilon^{ij}\vec S \cdot \partial_i \vec S
\times \partial_j \vec S.\label{charge}
\end{equation}
The topological charge density $q(\vec r)$ characterizes the spatial
distribution of the Skyrmion. The total topological charge or the
Pontryagian index $Q\equiv \int d\vec r q(\vec r)$ is an
invariant.\cite{Girvin,Moon} From the right panel of Fig. 5, one
finds that the singly quantized vortices have the Dirac
$\delta$-function topological charge density. On the other hand, the
giant Skyrmion has its charge uniformly distribute on a ring where
the two imiscable condensates overlap.

\section{discussions}
The additional rotation angle $\Delta \phi$ can be accounted for by
parameterizing the wavefunction as
\begin{equation}
\left(\begin{array}{c}\chi_1\\\chi_2\end{array}\right)
=\left(\begin{array}{c}e^{i\theta_1}\cos\beta(r)/2\\
e^{i\theta_2}\sin\beta(r)/2\end{array}\right).\label{para}
\end{equation}
To simplify the problem, we omit the chain of singly quantized
vortices and focus our attention on the giant vortex. The
configuration satisfies the boundary condition $\beta(0)=0$ and
$\beta(\infty)=\pi$, which is referred as the Anderson-Toulouse
vortex. The phase of majority condensate is constant which can be
set as $\theta_1=0$. The phase of minority condensate is independent
of radius $r$ and can be approximately written as $\theta_2=m\phi$.
Hence the spatial rotation of an angle $2\pi/k$ corresponding to a
relative phase difference $\Delta \theta=2\pi m/k$.

The pseudo-spin vector $\vec S$ can be expressed in the polar
coordinates
\begin{equation}
\vec
S=\{\sqrt{1-[S_z(r)]^2}\cos(m\phi),\sqrt{1-[S_z(r)]^2}\sin(m\phi),S_z\}.
\end{equation}
The topological charge density of Eq.[\ref{charge}] has the form
\begin{equation}
q(\vec r)=\frac{m}{4\pi r}\frac{dS_z(r)}{dr},
\end{equation}
which leads to the total charge of the giant Skyrmion as
\begin{equation}
Q=\int d^2\vec r q(\vec r)=\frac{m}{2}[S_z(\infty)-S_z(0)]=m.
\end{equation}

In summary, we have created a giant vortex in a two-species Bose
condensate with unequal particle numbers. The corresponding
pseudo-spin texture shows a $m$-fold symmetry and the topological
charge is $m$.

{\bf Acknowledgement} This work is supported by the National Natural
Science Foundation of China under grant No. 10574012.

\end{document}